\documentclass[aps,prd,twocolumn,superscriptaddress,longbibliography,nofootinbib]{revtex4-2}

\usepackage[utf8]{inputenc}
\usepackage{epigraph}
\usepackage{amsmath}
\usepackage{wrapfig}
\usepackage{epsfig}
\usepackage{amssymb}
\usepackage{graphicx}
\usepackage{slashed}
\usepackage{xcolor}
\usepackage{comment}
\usepackage{natbib}
\usepackage{enumitem}
\usepackage{tikz-cd,tikz}
\usepackage{float}
\usepackage{array,adjustbox,booktabs}
\usepackage{subcaption}
\usepackage{amsfonts}
\usepackage{physics}
\usepackage{hyperref}

\usepackage{xcolor}
\usepackage{colortbl}
\definecolor{dark-red}{rgb}{0.80,0.12,0.12} 
\definecolor{dark-blue}{rgb}{0,0.15,0.85} 

\hypersetup{
  linktocpage,
  colorlinks  = true, 
  urlcolor    = magenta, 
  linkcolor   = dark-blue, 
  citecolor   = purple 
}

\newcommand{\bea}{\begin{eqnarray}}
\newcommand{\eea}{\end{eqnarray}}

\usepackage[a4paper]{geometry}
\DeclareMathAlphabet\mathbfcal{OMS}{cmsy}{b}{n}

\usetikzlibrary{decorations.markings}
\tikzset{middlearrow/.style={
		decoration={markings,
			mark= at position 0.5 with {\arrow{#1}} ,
		},
		postaction={decorate}
	}
}

\def\be{\begin{equation}}
\def\ee{\end{equation}}
\def\eqn#1{eq.~\eqref{#1}}

\def\rcites#1{refs.~\cite{#1}}

\def\redA{\hat{\mathcal{A}}}

\def\intsum{\mathop{\sum \kern-1.2em \int}\limits_{\check{s}}d \hat{s}}

\def\intsumuno{\mathop{\sum \kern-1.2em \int}\limits_{\sum_j\!\check{s}_{ij}=\,p_i-1}\kern-1.5em{d} \hat{s}}
\def\intsumdue{\mathop{\sum \kern-1.2em \int}\limits_{\sum_j\!\check{s}_{ij}=\,p_i-2}\kern-1.5em{d} \hat{s}}
\def\intsumtre{\mathop{\sum \kern-1.2em \int}\limits_{\sum_j\!\check{s}_{ij}=\,p_i-3}\kern-1.5em{d} \hat{s}}

\def\oneloop{\begin{tikzpicture}[scale=0.06]
\def\shift{0.1};
\def\shiftt{0.2};
\draw[thick,dashed] (-1.2,-0.5)--(0.33,0.6);
\draw[thick,dashed] (-1.2,2.5)--(0.33,1.6);
\draw[thick,dashed] (4.5-3,0.5)--(6.3-3,-0.6);
\draw[thick,dashed] (4.5-3,1.6)--(6.3-3,2.5);
\fill[white] (1,1) circle (1);
\draw[thick] (1,1) circle (1);
\end{tikzpicture}
}

\newcommand{\nocontentsline}[3]{}
\let\origcontentsline\addcontentsline
\newcommand\stoptoc{\let\addcontentsline\nocontentsline}
\newcommand\resumetoc{\let\addcontentsline\origcontentsline}

\begin{document}

\title{Anomaly cancellation and  one-loop finiteness of 6D half-maximal supergravities}
\author{Renata Kallosh}
\email{kallosh@stanford.edu}
\affiliation{Leinweber Institute for Theoretical Physics at Stanford, 382 Via Pueblo, Stanford, CA 94305, USA}

\begin{abstract}
\noindent 

We suggest that the surprising one-loop finiteness of 6D half-maximal supergravities recently discovered in \cite{Huang:2025nyr} might be related to anomaly-free $SO(5, 21)$ G-duality symmetry in $(2,0)$ supergravity with 21 tensor multiplets and to anomaly-free $SO(4, 20)$ G-duality symmetry of $(1,1)$ supergravity with 20 vector multiplets. Our argument is reminiscent  of the analogous role of the  anomaly-free $E_{7(7)}$ G-duality symmetry in the UV finiteness of a maximal 4D supergravity at 3  loops.
\end{abstract}

\maketitle

\stoptoc

\parskip 6.3pt

\section{Introduction}
The surprising discovery of one-loop finiteness in 6D half-maximal supergravities in \cite{Huang:2025nyr} raises a few issues. 

1. How can the presence of specific matter improve the UV finiteness? In 4D in pure supergravities without matter with account of the Gauss-Bonnet theorem, the relevant dimension 4 counterterms for 1-loop  are given by $(R_{\mu\nu})^2$ and $R^2$ \cite{tHooft:1974toh}. This makes pure gravity 1-loop UV-finite on-shell. However, in the presence of matter when $R_{\mu\nu}$ does not vanish on shell but depends on the energy-momentum tensor, the counterterms become equal to $(T_{\mu\nu})^2$ and $T^2$, and the matter-coupled gravities generically have a 1-loop UV divergence.  Analogous analysis in terms of superfields is available in supergravity. Examples of 1-loop UV divergent supergravities coupled to matter supermultiplets are known \cite{Deser:1974cz,Fischler:1979yk}.

In 6D in pure gravity, the linearized counterterm has dimension 6, which allows a term like $R^3_{\mu\nu\lambda\delta}$. But this one has no supersymmetric extension and is excluded. The one that has a supersymmetric extension is a local counterterm 
\be
\int d^6 x (R_{\mu\nu\lambda \delta})^4
\label{R4}\ee
But it has dimension  $-6+8=2$   and therefore needs a factor $\Box ^{-1}$ which will make it a non-local expression. Therefore, it is not a valid 1-loop counterterm for a 4gr amplitude.  This is a simple reason why the 1-loop 4-graviton amplitude is finite in supergravity, in both the (2,0) and (1,1) cases.

Thus, in pure 6D supergravity, the 1-loop  4-particle amplitude is UV finite. Once matter multiplets are added, 1-loop UV divergences are expected, in general, so the result in  \cite{Huang:2025nyr} is indeed unexpected. We will present below the valid BPS candidate counterterms involving matter multiplets.

2. The   (2,0) supergravity model with 21 tensor multiplets is known to have a vanishing gravitational anomaly and to follow from type IIB  ten-dimensional chiral supergravity theory compactified on $K_3$ \cite{Townsend:1983xt}. We now know that this theory is 1-loop UV finite. Why? Is the cancellation of gravitational anomalies the reason for it? 

3. It is known that both   (2,0) supergravity model with 21 tensor multiplets and (1,1) supergravity with 20 vector multiplets originate from type II string theories compactified on $K_3$~\cite{Aspinwall:1996mn}. The low-energy limit of type IIB yields $\mathcal{N}=(2,0)$ supergravity with 21 tensor multiplets, while type IIA  (or heterotic string compactified on $T^{4}$) yields $\mathcal{N}=(1,1)$ supergravity with 20 vector multiplets. These theories are related by string dualities. 

{\it All facts about these theories have been known for decades, but it was never predicted that they are UV finite at 1-loop.} We have learned it only\footnote{It remained unnoticed that in the (1,1) case the 1-loop amplitude computation was performed in \cite{Bern:2013qca}.} from the actual amplitude computations  \cite{Huang:2025nyr}. 

Supergravities with scalar fields in the ${G\over H}$ coset spaces have {\it  hidden symmetries}. The hidden  
  $G$ symmetry is a global non-compact on-shell symmetry of supergravity, whereas  $H$ is a local composite symmetry of the action, $H$ being a maximal subgroup of $G$. In cases studied in \cite{Huang:2025nyr} both models have
\be
{G\over H}= {SO(p, q)\over SO(p) \times SO(q)}
\ee
Namely  
 in (2,0) 6D supergravity with $n_T$ tensor multiplets  we have \cite{Romans:1986er,Riccioni:1997np} 
\bea
&&{G\over H}\Big |_{(2,0)}= {SO(5, n_T)\over SO(5) \times SO(n_T)} \ .
\label{ten}\eea
In (1,1) 6D supergravity with $n_V$ vector multiplets (F(4) supergravity  coupled to matter),   there is the following coset space with one physical scalar in supergravity multiplet and $4n_V$ scalars in matter multiplets \cite{Romans:1985tw,DAuria:2000afl,Andrianopoli:2001rs}: 
\bea
&& {G\over H}\Big |_{(1,1)}= {SO(4, n_V)\over  SO(4) \times SO(n_V)}\times O(1,1) \ .
\label{vec}\eea
The physical scalars are in the corresponding coset spaces after local $H$ symmetry is gauge-fixed.  Both of these coset spaces in   \eqn{ten} and  \eqn{vec}  in case $n_T=21$ and $n_V=20$ were also identified in \cite{Aspinwall:1996mn} via compactification of string theory on $K_3$.
A recent survey of supergravities in 
 \cite{Sezgin:2023hkc} has a nice presentation of various 6D supergravities.

The  surprising discovery in   \cite{Huang:2025nyr} reminds the discovery in  \cite{Bern:2007hh}  that the 3-loop UV divergence is absent in 4D maximal supergravity with  
\be
{G\over H}={E_{7(7)}\over SU(8)}
\ee
 coset space.  This duality $G$ symmetry $E_{7(7)}$ in 4D  does affect the UV properties of supergravity. See, for example, \cite{Beisert:2010jx,Freedman:2018mrv} where it is predicted, assuming $E_{7(7)}$ is anomaly-free,  that maximal 4D supergravity is UV finite up to 6 loops.  In any case,  almost 2 decades after the discovery in  \cite{Bern:2007hh}, $E_{7(7)}$  remains the only known explanation of the 3-loop UV finiteness.
 
 The $E_{7(7)}$ precedent suggests that one should pay close attention to {\it the hidden $G$ symmetries of supergravities} when looking at the UV behavior of amplitudes. We will show, indeed,  that both cases of  1-loop  UV finite 6D theories coupled to matter might be related  to 
  $G=SO(p,q)$ symmetries, assuming these hidden G-symmetries and supersymmetries are anomaly-free, as in the case of $E_{7(7)}$ in maximal 4D supergravity.

 \section{Dualities as hidden $G$ symmetries of supergravity}
 
In the context of superamplitudes the main difference between G-dualities $E_{7(7)}$ and $SO(p,q)$ is the fact that in $E_{7(7)}$ case only a soft scalar limit of 6-point amplitudes was investigated\footnote{The $E_{7(7)}$ invariants are complicated, for example there is no metric,  quadratic invariants are antisymmetric.}    
 whereas in orthogonal groups, one can easily study both  4-point $SO(p,q)$ invariants as well as a soft scalar limit of 6-point amplitudes.

The nature of the orthogonal symmetry group $SO(p,q)$ can be best understood by the {\it Lorentz group $SO(1,3) $ analogy}. We build the invariants from the vector representation $x^\mu=(0,1,2,3)$ using  the metric $g_{\mu\nu}: + - - -$ with 1 time direction and 3 space directions. For example there are 4-point invariants $(x^\mu x^\nu g_{\mu\nu})^2= \left ((x^0)^2-\vec x)^2\right)^2$. If $(x^0)^4$ is absent, Lorentz symmetry predicts that $(x^0)^2\times (\vec x)^2$ and $\left((\vec x)^2\times (\vec x)^2) \right )$ are also absent.

In  $SO(5, n_T)$ we have 5 ``time'' directions and $n_T$ ``space'' directions and $\mu$ takes $5+n_T$ values.  In $SO(4, n_V)$ we have 4 ``time'' directions and $n_V$ ``space'' directions and $\mu$ takes $4+n_V$ values. In both cases, we have  4-point invariants for the relevant metric $(x^\mu x^\nu g_{\mu\nu})^2$.

The  4gr amplitudes in  $(2,0)$ and $(1, 1)$ supergravities are UV finite since there are no valid candidate counterterms. It means  that the ``time'' components in quartic invariants  $(x^\mu x^\nu g_{\mu\nu})^2$ are absent.  However, the  ``time-space'' 2gr2mat and ``space-space'' 4mat amplitudes have valid candidate counterterms for any $n_T$, $n_V$. We will show the counterterms explicitly below. The computations in   \cite{Huang:2025nyr} established that the valid candidate counterterms did not show up as UV infinities in the case of $n_T=21$, $n_V=20$. The $SO(5, 21)$ invariant  is
\be
(x^\mu x^\nu g_{\mu\nu})^2= \left (\sum_{\mu=0}^{\mu=4}(x^\mu)^2- (\sum_{\mu=5}^{\mu=26} x^\mu)^2\right)^2
\label{20}\ee 
The $SO(4, 20)$ invariant is
\be
(x^\mu x^\nu g_{\mu\nu})^2= \left (\sum_{\mu=0}^{\mu=3}(x^\mu)^2- (\sum_{\mu=4}^{\mu=24} x^\mu)^2\right)^2
\label{11}\ee  
In view of the vanishing 4gr  UV infinity  $\left (\sum_{\mu=0}^{\mu=4}(x^\mu)^2\right)^2$  the absence of UV infinities for 2gr2mat $\sum_{\mu=0}^{\mu=4}(x^\mu)^2  \times \sum_{\mu=5}^{\mu=21} x^\mu)^2$ and 4mat $\sum_{\mu=5}^{\mu=21}(x^\mu)^2  \times \sum_{\mu=5}^{\mu=21} x^\mu)^2$, despite the existence of the valid counterterms,  may be interpreted as a consequence of an anomaly-free $SO(5, 21)$ given in \eqn{20}.
Analogous explanation is valid for  (1,1) supergravity based on \eqn{11}.

Note that the $G=SO(p,q)$ symmetries act on 5 tensors in the gravitational multiplet and $n_T$ tensors in the matter multiplet in the (2,0) case, or on 4 vectors in the gravitational multiplet and $n_V$ vectors in the matter multiplet in the (1,1) case. To transfer the features of 4-tensor or 4-vector amplitudes to total superamplitudes in gravity multiplet and matter multiplet, including amplitudes of various spin fields, we have to study more carefully the {\it supersymmetry and duality}  issues in the context of UV infinities. 

In particular, we need the anomaly-free G symmetries as well as anomaly-free supersymmetry for our explanation of the computations in  \cite{Huang:2025nyr} to be valid.

\section{(2,0) supergravity and $SO(5,n_T)$ duality}
The generalized supergravity dualities in 6D were introduced by Romans \cite{Romans:1986er}, where the system of $p$ self-dual and $q$ anti-self-dual tensor fields has a natural $SO( p, q)$ group of duality transformations.

The chiral (2,0) supergravity with ${G\over  H}= {SO(5, n_T)\over SO(5) \times SO(n_T)}$ \cite{Townsend:1983xt,Romans:1986er,Riccioni:1997np} in gravitational multiplet has gravitini and the two-form potential field strength's is chiral, but in tensor matter multiplets the fermions and two-form potential field strength's have opposite chirality. The gravitational anomaly of this model cancels for $n_T=21$ \cite{Townsend:1983xt}. There is a relation between gravitational and supersymmetry anomalies due to the algebra of two supersymmetries, where
$
[\delta_1, \delta_2]= \delta_{gct} + \dots
$
A necessary condition for the absence of the supersymmetry anomaly is the absence of the gravitational anomaly. Therefore, it is reasonable to use supersymmetry for the analysis of counterterms. 

6D  (2,0) supergravity   fields are, in notation of \cite{Riccioni:1997np}: the vielbein $e_\mu^a$,  $a$ is a Lorentz index, a left-handed gravitino $\psi_\mu$, $(n+5)$ antisymmetric tensors $B_{\mu\nu}^r$ $ (r = 0, ..., n+4)$ obeying (anti)self-duality conditions, $n$ right-handed tensorini $\chi^m$  $(m = 1, ..., n)$ and $5n$ scalars. The  physical scalars parametrize the coset space ${SO(5, n)\over SO(5)\times SO(n)}$, and are associated to the $SO(5, n)$ matrix
\be
V= \left(\begin{array}{c}v_r^i \\x_r^m\end{array}\right) \ ,
\ee
$i=1,\dots, 5$, $m=1, \dots , n$. For example, for $n=21$, the number of scalars before local $SO(5)\times SO(21)$ gauge symmetries is fixed at 325, but there are only 105 physical scalars. The $SO(5, n)$ constraints on scalars are 
\bea\label{metric}
&&v^{ir} v_r^i=\delta^{ij}, \quad 
x^{mr} x_r^n=-\delta^{mn}, \\
&& v^{ir} x_r^m=0, \quad v_r^i v_s^i - x_r^m x_s^m = \eta_{rs} \ .
\eea
The local $SO(5)$ and  $SO(n)$ symmetries  have composite connections $Q^{ij}_\mu = v_r^i(\partial_\mu v_r^j)$, $S_\mu^{mn}= (\partial_\mu x^m_r) x^{nr}$. 
 The 
$SO(5, n)$ duality of the theory relates antisymmetric tensors of the theory.
We have to look at $5+n$ antisymmetric tensors \cite{Romans:1986er,Riccioni:1997np} where their (anti) self-duality condition was introduced. To the lowest order in the fermi fields, it is
\be
G_{rs} H_{\mu\nu\rho}^s = {1\over 6e} \epsilon_{\mu\nu\rho \alpha \beta \gamma} H^{\alpha \beta \gamma}_r
\label{Ricc}\ee
with
$
G_{rs}= v_r^i v_s^i + x_r^m x_s^m \ .
$
These conditions mean  that  5 of $H^i= v^i_r H^r$  are self dual, while $n$ of $H^m= x^m_r H^r$ are anti-self-dual. The divergence of the (anti) self-duality condition \eqn{Ricc} is $D_\mu (G_{rs} H_{\mu\nu\rho}^s)=0$.

The $SO(5, n)$ symmetry is implemented into the action as a gauge-fixed theory 
Lorentz invariant action for chiral p forms
\cite{Pasti:1996vs}, so that the bosonic action is
\be
e^{-1} {\cal L}_{bos}=- {1\over 4} R +{1\over 12} G_{rs} H_{\mu\nu\rho}^r H^{s\, \mu\nu\rho} +e^{-1} {\cal L}_{bos}(v, x) \ .
\label{Pasti}\ee
The 1-loop  counterterm, in addition to being supersymmetric, has to be constructed in an $SO(5, n)$ invariant way and has to depend on $n$ anti-self-dual matter tensors  $H^m$ and 5 self-dual gravity tensors $H^i$, but only in $SO(5, n)$-invariant combination. 
At the linear level, we have an $5+ n$ vector $H_r$ 
\be
H_r = (V\cdot H)_r = v^i_r H^i + x^m_r H^m  \ .
\ee
and the metric has 5 pluses and $n$ minuses. The action in \eqn{Pasti} has a quadratic invariant $G_{rs} H^r H^s$. The linearized 4-point counterterm for 4 tensors has the form $(G_{rs} H^r H^s)^2$.
Therefore, the absence of UV divergence for 5 self-dual gravity tensors $(H^i)^4$ requires the absence of UV divergence for a mix of two self-dual gravity tensors and two  anti-self-dual gravity tensors $(H^i)^2(H^m)^2$ as well as four anti-self-dual gravity tensors $(H^m)^4$.  However, these formal requirements are valid only if $SO(5, n)$ has no anomalies. 

\subsection{BPS candidate counterterms}

Explicit local supersymmetry transformations in \cite{Riccioni:1997np} allow us to build the linearized on-shell superfield for each multiplet. Supersymmetry parameters are $Sp(4)$ spinors with account of the isometry of  Lie algebras of $Sp(4, C)$ and $SO(5,C)$.

The on-shell linearized gravitational multiplet starts with 5 tensor field strength $H_{\mu\nu\lambda}^i$, which transforms into the gravitino field strength, with the next component being a space-time curvature, and after that the derivatives of these 3 types of fields. The superfield $H_{\mu\nu\rho}^i$ has dimension 1, as one can see from the action in eq. (2.10) in \cite{Riccioni:1997np}. 
It is a 1/2 BPS multiplet in which the manifest $Sp(4)$ symmetry is broken to $SU(2)$. It can be given in a form with two spinorial $Sp(4)$ indices $a=1,  \dots, 4$
\be
(H_{\mu\nu\lambda})_{ab}= \Gamma^i_{ab}H_{\mu\nu\rho}^i
\ee
The matter $n_T$  multiplets of dimension 0 start with a scalar,  the second component of the matter superfield is a spin 1/2 field, and the third one is the tensor field strength $H_{\mu\nu\lambda}^m$. The on-shell linearized matter  multiplet starts with $m$ scalars, also in a form with two spinorial $Sp(4)$ indices   
\be
W^m_{ab}
\ee
It has a dimension of 0. The second component of the matter superfield is a matter spin 1/2 field, and the next one is the tensor field strength $H_{\mu\nu\lambda}^m$.
Both multiplets
\be
\Phi_{ab} =\{ (H_{\mu\nu\lambda})_{ab}, \, W^m_{ab} \}
\ee
are BPS multiplets, as described in \cite{Howe:1983fr,Kallosh:2023dpr,Kallosh:2024lsl}, namely
\be
D_a \Phi_{bc} = \Omega_{ab} \Psi_c + \Omega_{ac} \Psi_b
\ee
We choose the following non-vanishing components of $ \Omega_{ab}$
\be
\Omega_{13}= \Omega_{24} =1
\ee
which means that the superfields $\Phi_{12}$ depend only on $\theta^3, \theta^4$ since $D_1\Phi_{12}=D_2\Phi_{12}=0$.

The BPS counterterm for the one-loop order 4-point amplitudes has the form
\be
\int  d^6 x (D_3 D_4)^4   \Phi_{12}^2 \Box^{n}  \Phi_{12}^2
\label{4p}\ee
Here $n$ takes values $-1, 0, 1$, see below. Only the cases with $n=0,1$ are local expressions. 
The part with 4 gravity multiplets $H^4$ is excluded by dimension $-6+4 +4=+2$ unless we make it non-local by inserting  $\Box^{-1}$
\be
I_{4gr}\to  \int  d^6 x (D_3 D_4)^4   H_{12}^2 \Box^{-1} H_{12}^2 \ ,
\ee
so it is not a valid 1-loop counterterm
The part with 2 gravity and 2 matter superfields has a correct dimension $-6+4+2=0$ and is a valid 1-loop counterterm
\be
I_{2gr2mat} \to \int  d^6 x (D_3 D_4)^4   H_{12}^2 W_{12}^2
\ee
Finally, the 4-matter multiplet candidate counterterm is a valid 1-loop counterterm  but requires an extra 2 derivatives so that we have $-6+4+2=0$
\be
I_{4mat} \to \int  d^6 x (D_3 D_4)^4    W_{12}^2 \Box W_{12}^2
\ee
Here we have $p=5$ fields $H$ and $q=m=n_T$ fields $W$. The statement that counterterms have $0\times I_{4gr}$  and non-zero for $I_{2gr2mat}$
and $I_{4mat}$  (as we have for $n_T\neq 21$ ) is a statement that $SO(5,n_T)$ is broken unless $n_T=21$ where all  3 UV divergences vanish. We have, symbolically
\be
0\times I_{4gr} + (n_T-21) I_{2gr2mat}+ (n_T-21)I_{4mat}
\ee
This means that at $n_T=21$  all 4-point amplitudes are UV finite
\be
0\times (I_{4gr} + I_{2gr2mat}+ I_{4mat})
\ee
If $SO(5,21)$ symmetry is anomaly-free, it would require this condition based on the fact that 4gr amplitude is finite and 2gr2mat and 4mat are $SO(5,21)$ partners of 4gr amplitude.

\subsection{Soft limits}
In the $E_{7(7)}$  case in 4D, it is more difficult to study invariants, compared with the orthogonal groups $SO(p,q)$. Therefore, the main analysis of the consequences of $E_{7(7)}$  for the amplitudes was via the scalar soft limits \rcites{Beisert:2010jx,Freedman:2018mrv,Kallosh:2023dpr,Kallosh:2024lsl}. The basic tool in these papers was to examine the 6-point amplitudes in the presence of 4-point UV divergences. There is a pole contribution to
the 6-point  matrix elements, which needs to be compensated in the soft scalar limit by a 6-point local counterterm. When such a local 6-point counterterm is not available, it signals the breaking of the soft limit. In maximal 4D supergravity, therefore, according to \rcites{Beisert:2010jx,Freedman:2018mrv} if $E_{7(7)}$  is anomaly-free, it protects the theory from UV divergences up to six loops: the relevant 6-point local superinvariants are not available.

In 6D a related analysis was performed in \cite{Kallosh:2023dpr,Kallosh:2024lsl} in (2,2) supergravity with $Sp(4)_L\times Sp(4)_R$ R symmetry broken down to $SU(2)\times SU(2)$ subgroup of the R symmetry. We can apply it to the case (2,0) supergravity with 1/2 BPS multiplet when the manifest $Sp(4)_L$ symmetry is broken down to $SU(2)$, the $Sp(4)_R$ symmetry is absent. First, we show that the 4-point BPS superinvariants, despite their  $Sp(4)$ symmetry, are broken down to $SU(2)$, actually have an $Sp(4)$ symmetry.

The 4-point BPS invariant in \eqn{4p} can be presented in the symbolic form
\be
\int  d^6 x (D_a D_b  \epsilon^{abcd})^2  (D_e D_f  \epsilon^{efgh})^2 \Phi^2_{cd} \Box^{n}  \Phi^2_{gh}
\ee
which shows that it is an $Sp(4)$ invariant. The reason this is an $Sp(4)$ invariant is the fact that in  \eqn{4p} we have 4 of each indices, 1,2,3,4
repeating 4 times, so we can contract them with $Sp(4)$ $\epsilon^{abcd}$ 4 times.

The BPS counterterm for the one-loop order 6-point amplitudes has the form
\be
\int  d^6 x (D_3 D_4)^4   \Phi_{12}^2 \Box^{n}  \Phi_{12}^4
\label{6p}\ee
 There are more $1,2$ indices than $3,4$ indices; there is no way to contract these with $Sp(4)$ $\epsilon$'s, 
and therefore these 6-point BPS invariants cannot be made $Sp(4)$ invariant. Therefore they are not valid local 6-point counterterms for $n=0,1$ cases, and the relevant 4-point UV divergences at $n_T\neq 21$  might break the soft scalar limits and $SO(5,n_T)$ at $n_T\neq 21$.

Our analysis of the soft limits is in complete agreement with the analysis of the anomaly-free $SO(5,n_T)$ symmetry predictions.


The formula for the four-matter one-loop amplitude in \cite{Huang:2025nyr} demonstrates it all beautifully. They have found
\begin{equation}\label{oneloopamp}
    \redA_{TT}^{\oneloop} =  \delta^{f_1f_2}\delta^{f_3 f_4}\!\mathcal{F}_{s}^{(2,0)}  + \delta^{f_1f_4}\!\delta^{f_2 f_3} \mathcal{F}_{t}^{(2,0)}+\delta^{f_1f_3}\delta^{f_2 f_4}\!\mathcal{F}_{u}^{(2,0)} \ ,
\end{equation}
where
\begin{align}\label{Btu}
\mathcal{F}_{s}^{(2,0)} \! = -  \frac{1}{(4\pi)^3} \Big[& t^2 \mathcal{B}_{s,t} +  u^2 \mathcal{B}_{s,u}  + \frac{n_T-21}{12} s \log\big({-}\frac{s}{\mu^2}\big) \Big] \ .
\end{align}
It demonstrates that gravitational anomalies and UV divergences are related, as explained above. In addition, the 5 self-dual and 21 anti-self-dual tensors are partners in $SO(5, 21)$, which holds at the 1-loop quantum level. As a result, this case is 1-loop UV finite. For any other non-vanishing $n_T$, the $SO(5, n_T)$ on-shell symmetry, present at the classical level, can be interpreted as being anomalous at the quantum level, as it follows from  \eqn{Btu}.

The  $SO(5, 21)$ symmetry requires the absence of UV divergences in mixed gravity and matter tensors, and for matter tensors, so that the 4-tensor gravity amplitudes and matter amplitude tensors are on equal footing.
Absence of anomalies in supersymmetry requires the absence of gravitational anomalies. Therefore,  in the $n_T=21$  case, supersymmetry promotes the absence of UV divergences from tensors in gravitational and matter multiplets to the absence of UV divergences for total gravitational and matter multiplets, not just tensor fields.

\section{(1,1) supergravity with $n_V=20$}

The non-chiral (1,1)  supergravity \cite{Romans:1985tw,DAuria:2000afl,Andrianopoli:2001rs} with  ${G/ H}= {SO(4, n_V)\over  SO(4) \times SO(n_V)}$ interacting with vector multiplets, is not expected to have gravitational anomalies for any number of vector multiplets $n_V$.\footnote{This was suggested to me by  H. Johansson and Y.-t. Huang, based on the double copy amplitude method.} Indeed, looking directly at the field content in non-chiral (1,1) supergravity \cite{Romans:1985tw,DAuria:2000afl,Andrianopoli:2001rs} one can see that the fermions are symplectic-Majorana and the tensor field strength is not restricted by (anti)-self-duality conditions.  Thus, neither fermions nor tensors contribute to the gravitational anomaly with any $n_V$.
Therefore, in the case of an arbitrary $n_V$ in (1,1) theory, it is reasonable to use supersymmetry for the analysis of counterterms. 

A relation between anomalies, soft-scalar limits, candidate counterterms, and UV divergences has been studied previously, for example, in \rcites{Beisert:2010jx,Freedman:2018mrv,Kallosh:2023dpr,Kallosh:2024lsl}. Here, in view of the new 1-loop results in  \cite{Huang:2025nyr}, it would be interesting also to study the direct consequences of $SO(4,n_V)$ symmetries as well as 
soft-scalar limits associated with these symmetries,  and to see if these soft limits are violated.\footnote{ This was suggested to me in private communication by C. Wen.} This would shed more light on the relation between the absence of $SO(4,20)$ anomalies in these models and 1-loop UV finiteness discovered in  \cite{Huang:2025nyr}.

The (1,1) supergravity with $n_V$ matter multiplets is described in detail in \cite{Andrianopoli:2001rs}. The field content of the supergravity multiplet and of the matter multiplets is presented there together with the action and local supersymmetry transformations. This allows us to deduce the features of the linearized superfields: the gravitational and the matter ones.

There are 4 vectors in gravity multiplets and $n_V$ vectors in $n_V$  matter multiplets. They are combined in a vector field strength $F^\Lambda_{\mu\nu}$ where  $\Lambda$ takes $4+n_V$ values.  
   The classical $SO(4, 20)$ duality symmetry acting on vectors shows up in every appearance of the vector field strength $F^\Lambda_{\mu\nu}$ in the action in eq. (3.56)-(3.58) in \cite{Andrianopoli:2001rs}. It enters only only via  $SO(4, 20)$ vector $F ^\Lambda$. Therefore, any candidate counterterm respecting $SO(4, 20)$ will depend on $F^\Lambda$. This puts vectors in the gravity multiplet and vectors in the matter multiplet on equal footing, which leads to  an on-shell 
  $SO(4, n_V)$ duality. 
  
  We proceed here directly with the analysis of the BPS invariants, following the procedure in the (2,0) case above.

\subsection{BPS candidate counterterms}

In the (2,0) case, we had the  1/2 BPS multiplet when the manifest $Sp(4)$ symmetry is broken down to $SU(2)$. It was given in a form with two spinorial $Sp(4)$ indices $a=1,  \dots, 4$. The superfields for gravity and matter were  $\Phi_{ab} =\{ (H_{\mu\nu\lambda})_{ab}, \, W^m_{ab} \}$, respectively and dimensions of $H$ is 1, dimension of $W$ is 0.

In (1,1) case \cite{Andrianopoli:2001rs} we have the R-symmetry group  $SO(4) \sim SU(2)_L \otimes SU(2)_R$ with  indices  $A=1,2$, $\dot A=1,2$. There are two linearized superfields, a gravity one and a matter one. As in (2,0) case also here $F$ has dimension 1, $W$ has dimension 0
\be
\Phi_{A\dot A} =\{ (F_{\mu\nu})_{A\dot A}, \, W^I_{A\dot A} \}
\ee
The BPS conditions are  \cite{Howe:1983fr}
\be
D_1 \Phi_{1\dot A}= D_{\dot 1} \Phi_{A\dot 1}=0
\ee
and the superfield $\Phi_{1\dot 1} $ depends only on $\theta^2, \theta^{\dot 2}$. A 4-point BPS superinvariant is  
\be
\int  d^6 x (D_2 D_{\dot 2} )^4   \Phi_{1\dot 1}^2 \Box^{n}  \Phi_{1\dot 1}^2
\label{4pv}\ee
Here $n$ takes values $-1, 0, 1$, for $F^4$, $F^2 W^2$,  $W^4$, respectively. As in the (2,0) case, the one for gravity is non-local 
\be
\int  d^6 x (D_2 D_{\dot 2} )^4   F_{1\dot 1}^2 \Box^{-1} F_{1\dot 1}^2
\label{4grav}\ee
and therefore excluded for any $n_V$, but the mixed and all  matter are valid 1-loop local supersymmetric counterterms
\be
\int  d^6 x (D_2 D_{\dot 2} )^4   F_{1\dot 1}^2   (W^I_{1\dot 1})^2
\label{4pmix}\ee
\be
\int  d^6 x (D_2 D_{\dot 2} )^4   (W^I_{1\dot 1})^2  \Box (W^I_{1\dot 1})^2
\label{4pm}\ee
How does this counterterm analysis compare with computations? In  (1,1) 6D supergravity  interacting with  $n_V$ vector multiplets the 1-loop amplitude in  \cite{Huang:2025nyr} has  the same structure as in \eqn{oneloopamp}, with $\mathcal{F}_{s}^{(1,1)}$ 
\begin{align}\label{Btu11}
\mathcal{F}_{s}^{(1,1)} \!= - \frac{1}{(4\pi)^3} \Big[& t^2 \mathcal{B}_{s,t} +  u^2 \mathcal{B}_{s,u}  + \frac{n_V-20}{12} s \log\big({-}\frac{s}{\mu^2}\big) \Big] \ , 
\end{align}
with similar expressions for the other crossing channels. In this case, the $\log(-s/\mu^2)$ term vanishes when $n_V=20$.

Assuming that supersymmetry has no anomalies, we conclude that all gravitational multiplet amplitudes are UV finite. This is confirmed by the computation in \cite{Bern:2013qca,Bern:2012gh}. We also find \cite{Bern:2013qca} that the matter 4-point amplitudes and the mixed ones have the features in  \eqn{Btu11}, which is the result in  \cite{Huang:2025nyr}: all 1-loop amplitudes are finite under the condition that $n_V=20$. 

Thus,  there is a UV finite $F^4$ amplitude and also UV finite $F^2 W^2$  and $ W^4$  amplitudes at $n_V=20$.  One can therefore interpret the result in \eqn{Btu11} and an analogous one in \cite{Bern:2013qca} as an absence of anomalies in the $SO(4, 20)$ symmetry at 1-loop order at $n_V=20$.  Supersymmetry, which is also assumed to be non-anomalous, enforces that the total matter multiplet and mixed ones' UV divergence vanish, not just their 4-vector part, which is controlled by the $SO(4, 20)$ symmetry.

For $n_V\neq 20$ we see in \eqn{Btu11} that $SO(4, n_V)$ symmetry at 1-loop order is broken since the supergravity multiplet is UV finite but matter and mixed amplitudes are UV infinite. So, the classical $SO(4, n_V)$ symmetry is anomalous at the quantum level at $n_V\neq 20$.

\subsection{Soft limits}
The 4-point BPS invariant \eqn{4pv} can be presented in the symbolic form
\be
\int  d^6 x (D_A D_{\dot A}  D_C D_{\dot C}  \epsilon^{AB} \epsilon^{\dot A \dot B}    \epsilon^{CD}\epsilon^{\dot C \dot D}  \Phi_{B\dot B} \Box^{n/2}  \Phi_{D\dot D})^2
\label{4pvS}\ee
which shows that it is an $SU(2)_L \otimes SU(2)_R$ invariant. Only $n=0,1$ correspond to local superinvariants. The reason this is an R-symmetry invariant is that in eq. \eqn{4pv} we have 2 of each indices, 1,2 and $\dot 1, \dot 2 $
repeating 4 times, so we can contract them with $\epsilon^{A\dot A}$ 4 times.

The BPS counterterm for the one-loop order 6-point amplitudes has the form
\be
\int  d^6 x (D_2 D_{\dot 2} )^4   \Phi_{1\dot 1}^2 \Box^{n}  \Phi_{1\dot 1}^4
\ee
This time we have more indices $1, \dot 1$ than $2, \dot 2$.  There is no way to contract these with $\epsilon^{A\dot A}$'s
and therefore these 6-point BPS invariants cannot be made $SU(2)_L \otimes SU(2)_R$  invariant. Therefore, they are not valid local 6-point counterterms for $n=0,1$ cases, and the relevant 4-point UV divergences at $n_V\neq 20$  break the soft scalar limits and $SO(4,n_V)$ at $n_V\neq 20$.

Thus, the soft limit analysis confirms the direct prediction from $SO(4, n_V)$ symmetry at 1-loop order.
At $n_V\neq 20$ $F^4$  amplitude is still UV finite, but  $F^2 W^2$ and   $W^4$ amplitudes are UV divergent.  That means that  $SO(4, n_V)$ symmetry at 1-loop order at $n_V\neq 20$ is broken; it is therefore an anomalous on-shell symmetry valid only at the tree level.

\section{Discussion}
We suggested hidden G-symmetries as a plausible underlying reason for the {\it one-loop UV finiteness} discovered in \cite{Huang:2025nyr}.  Our suggestion is based on the following observations.
 
 1) Absence of 1-loop counterterm in pure (2,0)  and  (1,1) 6D supergravity. The reason for this is dimension and supersymmetry.
 In the (1,1) case, the 1-loop UV finiteness was also confirmed directly in computations in \cite{Bern:2012gh} and in \cite{Bern:2013qca}.

 2) We presented here the 4-point candidate 1-loop counterterms in (2,0)  and in (1,1) supergravities and confirmed that the 4-point gravitational multiplet does not have valid 1-loop candidate counterterms.  Meanwhile, the  4-point 
 matter and mixed gravity-matter amplitudes have valid candidate BPS counterterms that preserve the R-symmetry, $Sp(4)$, and $SU(2)_L \otimes SU(2)_R$ in (2,0) and (1,1) supergravities, respectively. 
 
3) Our main concern therefore  is the {\it absence of anomalies in $G$-symmetries},  $SO(5, 21)$ in (2,0) and  $SO(4, n_V)$ in (1,1) case.
  $G$ duality symmetries  $SO(5, 21)$   and $SO(4, 20)$  force matter tensors or vectors to be partners of tensors or vectors of the gravitational multiplet. These are $SO(p,q)$ partners of the same nature as time and space coordinates $x^\mu$ in 4D where $\mu=0,1,2,3$ with the metric 
 $+ - - -$. Here in (2,0) 5 tensors in gravity and 21 in matter form a tensor $H^r_{\mu\nu\lambda}$ with $r$ taking $5+21$ values and metric with 5 pluses and 21 minuses, see \eqn{metric}. In (1,1) case 4 vectors in gravity and 20 in matter form a vector  $F^\Lambda_{\mu\nu}$ with $\Lambda$ taking $4+20$ values and metric with 4 pluses and 20 minuses.

 Gravitational tensors (vectors) according to 1) above are free of 1-loop divergences; therefore, matter tensors (vectors) must also be free of 1-loop divergences due to $SO(5, 21)$  ($SO(4, 20)$). If {\it supersymmetry is anomaly-free}, the UV finiteness of 4 tensors or vectors transfers to UV finiteness of all 4-point superamplitudes.
 This clarifies in the 1-loop case a relation between the absence of anomalies in $SO(5, 21)$  ($SO(4, 20)$), and the properties of UV divergences.

{\it The second issue is whether the UV finiteness in these models extends beyond one loop}. As we mentioned in point 1) above, the absence of the 1-loop counterterm in pure (2,0)  and  (1,1) 6D supergravity is accidental, due to dimension and supersymmetry. Points 2), 3), and 4) are valid only if point 1) is valid. This means that all amplitudes with matter superfields involved have to be UV finite based on anomaly-free $SO(5, 21)$   and $SO(4, 20)$ symmetries, despite the existence of the relevant counterterms.

At the 2-loop order, one expects  UV infinities in pure (2,0)  and  (1,1) 6D supergravity. The difference from 1-loop in the 4-graviton sector is the existence of the invariant depending on 4 Riemann-Christoffel curvature tensors
$
\kappa^2\int d^6x R^2 \Box R^2
$.
In pure (2,0), the candidate counterterms involving 4-point gravitons could be of the form
$\kappa^2 \int  d^6 x d^8 \theta H^2 \Box H^2(x, \theta)$ and they have a correct dimension 
 $-4-6+4+2= 0$.   Analogous counterterms are available in 2-loop pure (1,1) supergravity.

In (1,1) supergravity, the actual 2-loop computation was performed in \cite{Bern:2013qca} and demonstrated the UV divergence in 4gr, mixed 2gra2mat vectors and 4mat vector amplitudes. None of them vanish at $n_V=20$. 
In case of (1,1) supergravity + 20 vectors, the loop computations were performed in \cite{Bern:2013qca} in a setting different from  \cite{Huang:2025nyr}, assuming that matter multiplets come from dimensional reduction from dimension $D_s= D+n_V$ so that the answer in 6 dimensions corresponds to $D_s=6+n_V$. In this setting, they have found that the 1-loop UV divergences are proportional to $26-D_s= 20-n_V$. This agrees with the computations in \cite{Huang:2025nyr}, which were performed directly in 6D. At 2-loop order, the result in \cite{Bern:2013qca} in eqs. (5.16) and (5.17) suggest that the UV subdivergences of the order ${1\over \epsilon^2}$ cancel at $26-D_s= 20-n_V=0$, but the UV divergences of the order ${1\over \epsilon}$ do not cancel at $ n_V=20$.

 It remains to be seen if the direct computation in 6D will  confirm the 2-loop UV divergence found in  \cite{Bern:2013qca}. If the direct 6D results show UV finiteness, we will have to study  (1,1) 6D supergravity with vector multiplets described in superspace in \cite{DAuria:2000afl,Andrianopoli:2001rs} in the context of Free Differential Algebra, and investigate the predictions of this theory.

In view of these facts about (1,1) supergravity + 20 vectors, it is hard to expect that the 2-loop computations will reveal UV finiteness in (2,0) supergravity + 21 tensors, if the 1-loop finiteness in both theories has the common origin, namely anomaly-free $SO(5, 21)$   and $SO(4, 20)$ symmetries.

However, in the context of (2,0) supergravity with tensor multiplets, it was recently suggested \cite{Kallosh:2026roi} that the 1-loop UV finiteness might be due to an underlying superconformal theory that is anomaly-free for $n_T=21$. The (1,1) supergravity does not have a superconformal origin. Therefore, the  only known explanation of 1-loop UV finiteness is   anomaly-free $SO(4, 20)$ symmetry (like $E_{7(7)}$ symmetry for a maximal 4D supergravity at 3-loop order).

It remains to be seen whether the future 2-loop computation in 6D will support the superconformal conjecture of \cite{Kallosh:2026roi} by demonstrating 2-loop UV finiteness, or whether a UV divergence will be found and the explanation of the 1-loop UV finiteness might still be the anomaly-free $SO(5, 21)$ symmetry.

\section{Acknowledgments}

I am grateful to Yu-tin Huang, Henrik Johansson, Andrei Linde, Congkao Wen, and Yusuke Yamada for many enlightening discussions, and especially to Yu-tin Huang, Henrik Johansson, and Congkao Wen for the most recent communications. I am grateful  to Arkady Tseytlin for collaboration on the recent work \cite{Kallosh:2026roi}, which helped make sharp predictions for future computations in (2,0) and (1,1) theories.
This work is supported by Leinweber Institute for Theoretical Physics at Stanford and by NSF Grant PHY-2310429.


\twocolumngrid

\bibliography{lindekalloshrefs}
\end{document}